\documentclass[12pt]{article}
\usepackage{amssymb,graphicx}
\newcommand{\argmin}{\mathop{\mathrm{arg\,min}}}

\newcommand{\ER}{\ensuremath{\mathbb{R}}}

\begin{document}

\title{ Using complex surveys to estimate the $L_1$-median of a functional variable: application to electricity load curves}

\author{{\sc Mohamed Chaouch }$^1$ \quad and \quad {\sc Camelia Goga}$^2$ \\
$^1$ EDF Recherche \& D\'eveloppement, D\'epartement ICAME,\\ 
Clamart - France\\
$^2$Institut de Math\'ematiques de Bourgogne, Universit\'e de Bourgogne, \\
 DIJON - France \\
email : $mohamed.chaouch$@edf.fr, $camelia.goga$@u-bourgogne.fr}

\maketitle

\begin{abstract}
Mean profiles are widely used as indicators of the electricity consumption habits of customers.  Currently, in \'Electricit\'e De France (EDF), class load profiles are estimated using point-wise mean function. Unfortunately, it is well known that the mean is highly sensitive to the presence of outliers, such as one or more consumers with unusually high-levels of consumption. In this paper, we propose an alternative to the mean profile: the $L_1$-median profile which is more robust. When dealing with large datasets of functional data (load curves for example), survey sampling approaches are useful for estimating the median profile avoiding storing the whole data. We propose here estimators of the median trajectory using several sampling strategies and estimators. A comparison between them is illustrated by means of a test population. We develop a stratification based on the linearized variable which substantially improves the accuracy of the estimator compared to simple random sampling without replacement. We suggest also an improved estimator that takes into account auxiliary information. Some potential areas for future research are also highlighted. 
\end{abstract}





\section{Introduction }


The French electricity company, \'Electricit\'e De France (EDF), uses extensively the customer class load profiles in distribution network calculation.  Load profiles are also used to predict future loads in distribution network planning or  to estimate the daily load curve of a new customer. The customer data usually includes information about  the type of the electricity connection, the customer class, the consumption type  and some other additional information. The combination of the individual customer informations and the class load profiles allows us to estimate its load curve.

At EDF, the mean profile is used as an indicator of the electricity consumption of the customers.  Nevertheless, it is well known that the mean is highly sensitive to the presence of outliers, for instance consumers with high level of consumption. As Small (1990) states,  ''it suffices to have only one customer contaminating a data set and going off to infinity to send the mean curve to infinity as well. By contrast, at least 50\% of the data must be moved to infinity to force the median curve to do the same``.  More precisely, the median is robust to punctually extreme electricity consumptions of some customers and from a practical point of view, this can help to manage  the electricity supply. Moreover, in the context of the electricity open market, new customers join the EDF company while other ones leave it and it is important to know the amount of electricity demand. Since the load profiles are not known for new customers, it is more difficult to predict their impact on the global electricity demand. Based on individual customer information, new customers will belong to a specific class and will be allowed the synthetic profile that describes the consumption behavior of its class. In these situations, robust profiles should be used and this is why, we suggest using the median profile besides the mean curve as a robust indicator for analyzing the population of electricity load curves. 

The median of a sample of univariate observations is a natural and useful characteristic of central position. Multivariate data, on the other hand, have no natural ordering. There are several ways to generalize the univariate median to multivariate data and they all have their advantages and disadvantages (see Small, 1990 for a survey of multidimentional medians). First uses of the multivariate median were limited to two-dimentional vectors and were motivated mainly by problems of quantitative geography (namely, of centro-graphical analysis) which were dealt with by the U.S.A. Bureau of census in the late 19th and early 20th century.

We focus here on the $L_1$-median, also called the geometric or spatial median. Early work on the spatial median is due to  Hayford (1902) and Gini \& Galvani (1929) among others.  Its definition is a direct generalization of the real median proposed by Haldane (1948) and properties of the spatial median have been studied in details by Kemperman (1987). Iterative estimation algorithms have been developed by Gower (1974) and Vardi and Zhang (2000).




In the next few years, the French company  EDF intends to install over 30 millions electricity smart meters, in each firm and household.  A meter is an electronic device constructed for measuring the electricity consumption. These meters will be able to send individual electricity consumption measures on very fine time scales.  The new smart electricity meters will provide accurate and up-to-date electricity consumption data that can be used to model distribution network loads. In view of this new setting, the interest variables such as the consumption curve for example, may be considered as realizations of  functional variables  depending on a continuous time index $t$ that belongs to $[0, T]$ rather than multivariate vectors.   Kemperman (1987), Cadre (2001) and Gervini (2008) studied the properties of the median with functional data. We cite also the very recent work of Cardot \textit{et al.} (2011).  

Another important issue is data storage.  The amount of load data will be enormous when all or almost all of the customers have smart meters. Collecting, saving and analyzing all this information, would be very expensive. For example, if measures are taken every 10 minutes during one year and if we are interested in estimating the total electricity consumption for residential customers, the data storage is of about 100 terabytes. 



We suggest using survey sampling techniques in order to get  estimates of the median profile that are as accurate as possible at reasonable cost. The reader is referred  to Fuller (2009) for a very recent monograph on survey sampling theory. Nevertheless, the idea of selecting randomly a sample from a population of curves is rather new. Chiky \& H\'ebrail (2008) compare two approaches for estimating the mean curve. The first one consists in using signal compression methods for the whole population of curves and the second approach suggests taking a simple random sampling of the actual curves. Their conclusion is that  the results are better  in the latter situation even with  rather simple sampling designs.   Very recently, Cardot \textit{et al.} (2010) developed the estimation of  functional principal component analysis with survey data and Cardot and Josserand (2011)  studied the properties of the mean curve estimator with stratified sampling.  We cite also Chaouch and Goga (2010) who treated the estimation of geometric quantiles, the generalization of the spatial median, with survey data. As far as we know nothing has been done in the estimation of the $L_1$-median in a functional framework with survey data whereas it might have great practical interest. This is why,  we investigate in this paper the median curve estimator when several sampling designs and estimators are used.
It is worth mentioning that the results presented in this paper may be applied for other functional data which are not necessarily related to time as it was the case of the electricity data.  Nowadays,  functional data may arise in various other domains such as chemometrics or remote sensing  and then the functional response variables depend on index $t$ that may be a frequency and not necessary a time index.


The paper is structured as follows: section 2 gives the main results concerning the median curve estimation with survey data. A weighted estimator for the median curve is suggested and its asymptotic variance function is exhibited by means of the linearization technique developed by Deville (1999). A variance estimator is also proposed. 
Section 3 gives the estimation of the median curve and of its variance function for a firm population of $N=18902$  load electricity curves.  We consider several sampling designs: the simple random sampling without replacement, the systematic sampling,  the stratified sampling with optimal and proportional allocation, and the with replacement proportional-to-size design. In the case of the stratified sampling, we suggest using the k-means algorithm to construct homogeneous strata   with respect to the linearized variables. We illustrate through simulations that a substantial reduction compared to simple random sampling is obtained.  We adapt to the functional framework the selection of a sample when auxiliary information is used at the sampling stage as for the with replacement proportional-to-size design. Finally, we improve the Horvitz-Thompson estimator of the   functional  median by considering the poststratified estimator. 

\section{Functional Median in a Survey Sampling Setting}

Let us consider the finite population $U=\{1, \dots, N\}$ of size $N$ and a functional variable $\mathcal Y$ defined for each element $k$ of the population $U :$ $Y_k(t),$ for $t \in [0,T],$ with $T < \infty.$  Let $<\cdot,\cdot>,$ respectively  $||\cdot||,$ be the inner product, respectively the norm, defined on  $L^2[0,T]. $  
The empirical  median trajectory calculated from $Y_1, \ldots, Y_N$ is defined as  (Chaudhuri, 1996 and Gervini, 2008)

\begin{eqnarray}\label{defm}
m_N=\argmin_{y\in L^2[0,T]} \sum _{k=1}^N||Y_k - y||. 
\end{eqnarray}

\noindent Supposing that  $Y_k, $ for all $k=1,\ldots, N,$ are not concentrated on a straight line,  the median exists and is unique  (Kemperman, 1987) and is the solution of the following estimating equation,
 \begin{eqnarray}
 \sum_{k=1}^N\frac{Y_k - y}{||Y_k - y||}=0\label{medU}
 \end{eqnarray}
provided that $m_N\neq Y_k$ for all $k=1,\ldots, N. $

\noindent For $Y_1, \ldots, Y_N\in \ER^d,$ the median $m_N$ defined by the  formula (\ref{defm}) arises as a natural  generalization of the well-known characterization of the univariate median (Koenker and Basset, 1978),
$$
q=\arg\min_{\theta} \sum_{k=1}^N|Y_k-\theta|
$$
\noindent and it was called the \textit{spatial median} by Brown (1983) or  \textit{the $L_1$-median} by Small (1990). Weber (1909) considered $m_N$ as a solution to a problem in a \textit{location  theory} in which the $Y_1, \ldots, Y_N$ are the planar coordinates of $N$ customers, who are served by a company that wants to find an optimal location for its warehouse. 
  It is also known as the Fermat-Weber point. A geometrical interpretation of the median defined by (\ref{medU}) is that the centroid of the vectors 
$\displaystyle\frac{Y_k-m_N}{||Y_k-m_N||}$ is the origin in $\ER^d. $ With only three points and bidimensional data, the median  $m_N$ is known to be the Steiner point of the triangle $Y_1Y_2Y_3.$  The spatial median has also origins in the early work during the twelve census in the United Sates in 1900 concerned by finding the geographical center of the population over time. Hayford (1902) proposed the point-wise median as the geographical center but explicitly noted the drawback of the fact that the point-wise median depends on the choice of the orthogonal coordinates and it is not equivariant under orthogonal transformations.  Brown (1983) goes further with this idea and states that when dealing with spatial data where variables possess isometry and require statistical techniques that have rotational invariance, it is more appropriate to use a median that shares these properties. We recall that with observations $Y_1, \ldots, Y_N$ that lie in $\ER^d, $ the point-wise median is the $d$-dimensional vector of medians computed from the univariate components  and for functional variables, the empirical point-wise median is obtained if the $L^1$ norm is used in (\ref{defm}) instead of the $L^2$ norm, 
$$
\mbox{med}(t)=\argmin_{y(t)\in \ER} \sum _{k=1}^N|Y_k(t) - y(t)|, \quad \mbox{for all } t\in [0,T]. 
$$

\begin{figure}[h!]
\begin{center}
\includegraphics[height=7cm,width=10cm]{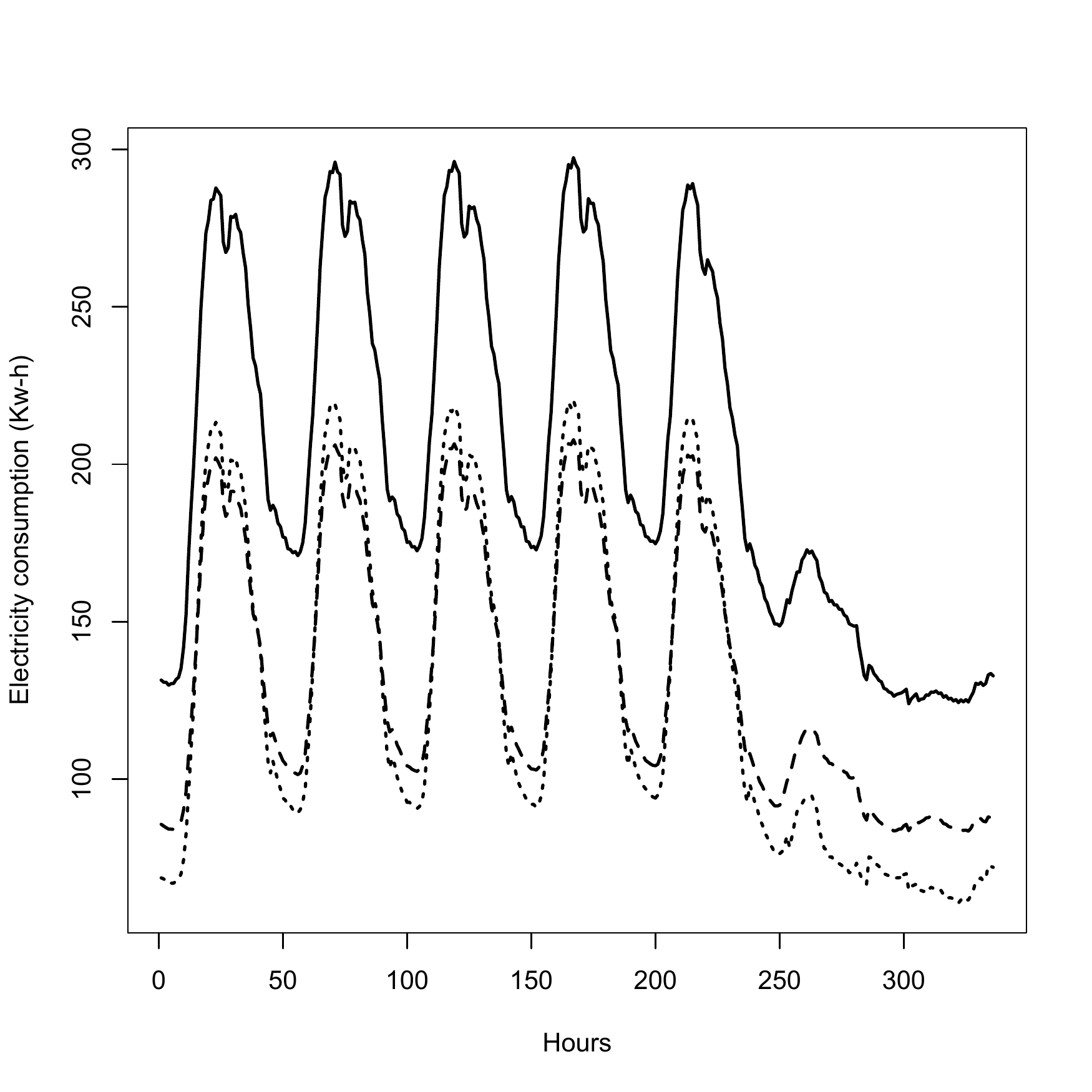} 
\end{center}
\caption{The spatial median profile is plotted in dashed line, the point-wise median profile in dotted line and the mean profile in solid line.}
\label{meanmed}
\end{figure}

To illustrate the mean curve and the point-wise median versus the spatial median, we plot   in Figure~\ref{meanmed} the three curves for the test population of $N=18902$ companies considered in section \ref{applicationedf}. The electricity consumption is measured every 30 minutes.

Chaudhuri (1996) shows that the geometric quantiles defined in formula (\ref{defmedgeom}) from below and which are a generalization of the median defined by (\ref{defm}) are equivariant under orthogonal transformations unlike the point-wise median.    Moreover, Chaudhuri showed also that the spatial or the $L_1$ median is equivariant under any homogeneous scale transformation of the coordinates of the multivariate observations which is appropriate when one needs to standardize the coordinate variables appropriately before computing the median. 

 The main arguments that play in favor of the spatial median  are the \textit{uniqueness} (see e.g. Chaudhuri, 1996 for the $d$-dimensional case with $d\geq 2$ and Kemperman, 1987 for the functional case) and the fact that it is \textit{a global and central indicator} of the distribution of the data. More exactly, the spatial median takes into account all instants making the spatial median  a central indicator of the distribution of the data while the point-wise median is a central indicator but only for each instant.  Consider for example that we have  consumption electricity data  recorded during two weeks: one working week and one holiday week such as the Christmas week. We compute first the spatial, respectively the point-wise median, by considering only the working week time measurements. Next, we consider the two week consumption electricity and  we compute again the spatial, respectively the point-wise median. It can be noticed that the coordinates of the point-wise median that correspond to the working week are the same in both situations while they are changed for the spatial median.  Since is due to the fact that the spatial median is computed by taking into account all the time measurements while the point-wise median is computed instant by instant.
 
Moreover, Brown (1983) shows that there is an asymptotic efficiency from using the spatial median instead of the point-wise median. In fact, one can see that the objective function is differentiable in the case of the spatial median while this property is not fulfilled for the point-wise median. 

As noted by Serfling (2002), the median defined by (\ref{medU}) and $Y_1, \ldots, Y_N\in \ER^d,$ has a nice robustness property in the sense that $m_N$ depends only on its direction towards $Y_i. $ More exactly, $m_N$ remains unchanged if the $Y_i$ are moved outward along these rays while it is obvious that the point-wise median will change. 



 
 \noindent\textbf{Remark}: Chaudhuri (1996) extends the definition (\ref{defm}) to geometric quantiles by using the geometry of data clouds. In a functional setting, its definition indexes the quantiles by the elements  $v\in L^2[0,T]$ with $||v||<1, $ 
  \begin{eqnarray}
 \mathcal Q(v)=\argmin_{y\in L^2[0,T]} \sum _{k=1}^N\left(||Y_k - y||+<Y_k - y,v>\right). \label{defmedgeom}
\end{eqnarray}
In this way, functional quantiles are characterized by a direction and magnitude specified by  $v\in L^2[0,T]$ with $||v||<1. $  Nevertheless, except the case $v=0,$ it is difficult to interpret the functional quantile defined in this way.  This is why, our discussion is limited to the case $v=0. $
\subsection{The design-based estimator for the median $m_N$}

Algorithms have been proposed to solve the equation (\ref{medU}) (Vardi and Zhang, 2000, Gervini, 2008) but they need important computational efforts especially when the number of time measurements is large. 
In this work, we suggest estimating the median curve $m_N$ by taking only a sample $s$ from $U$ according to a sampling design. A probability measure $p(\cdot)$ on the set of subsets of $U,$ henceforth denoted $\mathcal{P}(U)$, is called a {\it sampling design}. Any random variable $S$ with values in $\mathcal{P}(U)$ and distribution $p$, is called a {\it random sample} associated to the sampling design $p.$ Let $s$ be a realization of $S. $ For any $k \in U$, the inclusion probability of $k$ is given by 

$$\pi_k = \mathbb{P}(k\in S)=\sum_{k\in s} p(s),$$

\noindent where the sum is considered over all samples $s$ containing the individual $k.$
\noindent If $k\neq l$ are two elements of $U$, the second-order inclusion probability of $k$ and $l$ is given by

$$\pi_{kl} = \mathbb{P}(k,l\in S) = \sum_{k,l \in s} p(s),$$
\noindent where the sum is considered over all samples $s$ containing both $k$ and $l.$

\noindent In practice, a wide variety of selection schemes are used. We distinguish direct element sampling designs such as the simple random sampling without replacement (SRSWOR), stratified sampling (STRAT) or  proportional-to-size sampling designs (with or without replacement). Most of these designs are used extensively in practice.  However, such designs require having a sampling frame list identifying every population element which may be difficult, expensive or even impossible to realize. In order to avoid it, more complex designs such as cluster or multi-stage designs can then be used. This is for example appropriate when the population is widely distributed geographically or may occur in natural clusters. Using such designs saves money and human efforts but entails a loss of efficiency.  A detailed presentation of the survey sampling theory and many practical illustrations can be found in Korn and Graubard (1999), Lehtonen and Pahkinen (2004) and the reference book of S\"{a}rndal, Swensson \& Wretman (1992).

\noindent The median $m_N$ given by (\ref{medU}) is a nonlinear parameter of finite population totals defined by an implicit equation. In order to estimate $m_N, $ we use the functional substitution approach proposed by Deville (1999) for multivariate variables $\mathcal Y$ and extended to functional variables $\mathcal Y$ by Cardot \textit{et al.} (2010). Let $M$ be the discrete measure defined on $L^2[0,T]$ assigning the unity mass to each curve $Y_k$ with $k\in U$ and zero elsewhere, namely
$$
M=\sum_{k\in U}\delta_{Y_k},
$$
where $\delta_{Y_k}$ is the Dirac function in $Y_k. $ The total mass of $M$ is $N,$ the population size. Let $T$ be the functional with respect to $M$ and depending of $y$ as follows
\begin{eqnarray}
 T(M;y)=-\int \frac{\mathcal{Y}-y}{||\mathcal{Y}-y||}dM=-\sum_{k\in U}\frac{Y_k-y}{||Y_k-y||}.\label{fonctionnelle_neg}
\end{eqnarray}
Remark that $T$ defined in this way is the derivative  with respect to $y$  of the objective function given in (\ref{defm}). The median $m_N$ is then defined as an implicit functional with respect to $M,$ 
\begin{eqnarray}
T(M;m_N)=0 \label{Tmedian}
\end{eqnarray}
or equivalently, 
\begin{eqnarray}
\int \frac{\mathcal{Y}-m_N}{||\mathcal{Y}-m_N||}dM=0.\label{fonctionalmedian}
\end{eqnarray}

\noindent Let $\widehat M$ be a weighted estimator of $M$ based on the sample $s,$
\begin{eqnarray}
\widehat M=\sum_{k\in s}w_k\delta_{Y_k}=\sum_{k\in U}I_kw_k\delta_{Y_k},
\end{eqnarray}
where $I_k=\mathbf{1}_{\{k\in s\}}$ is the sample membership indicator of element $k\in U$ (S\"{a}rndal \textit{et al.}, 1992). In fact, $\widehat M$ is also a discrete and finite measure assigning the weight $w_k$ for each $Y_k$  with $k\in s$ and zero elsewhere.  Usually, one  take $w_k=1/\pi_k,$ the Horvitz-Thompson weights. In this case, we obtain the Horvitz-Thompson (1952) of $M$ which estimates unbiasedly the measure $M$ since $E_p(I_k)=\pi_k, $ for all $k\in U$ for $E_p(\cdot)$  the expectation with respect to the sampling design $p(\cdot).$ The reader is referred to Cardot \textit{et al.} (2010) and Cardot and Josserand (2011) for more details about the Horvitz-Thompson estimation with functional data. However, for $Y_1, \ldots, Y_N$ lying in $\ER^d,$ weights that take into account auxiliary information have been suggested. We mention Deville (1999) for  calibration weights or the very recent work of Goga and Ruiz-Gazen (2011) for nonparametric weights. Nevertheless, the extension to the functional case is not straightforward and it will be treated elsewhere.  For the rest of the paper, we consider $w_k=1/\pi_k$ and in section \ref{generalsettings}, we suggest the poststratified estimator of $M.$\\
\noindent Plugging $\widehat M$ into the functional expression of $m_N$ given by (\ref{Tmedian}), yields the design-based estimator $\widehat m_n$ of $m_N.$  Hence, $\widehat m_n$ verifies 
 $$
 T(\widehat M;\widehat m_n)=0,
 $$
namely, $\widehat m_n$ is the solution of the design-based estimating equation,
\begin{eqnarray}
\sum_{k\in s} \frac{1}{\pi_k} \frac{Y_k-\widehat m_n}{||Y_k-\widehat m_n||}=0.\label{medestim}
\end{eqnarray}
\noindent Supposing now that   for all $k\in s,$  $Y_k\neq \widehat m_n $  and that  $Y_k$ are not concentrated on a straight line,  we obtain that  the solution $\widehat m_n$ exists and is unique following the same arguments as in Kemperman (1987) or Chaudhuri (1996).   The median estimator $\widehat m_n$ is also called \textit{the substitution estimator} of $m_N$ and it is defined by a non-linear implicit function of Horvitz-Thompson estimators. As a consequence, the variance as well as the variance estimator of $\widehat m_n$ can not be obtained directly using Horvitz-Thomson formulas. We will give in the next a first-order expansion of $\widehat m_n$ in order to approximate $\widehat m_n$ by the Horvitz-Thompson estimator for the finite population total of appropriate artificial variables.

\subsection{Asymptotic properties}

The functional $T$ given by (\ref{fonctionnelle_neg})  is Fr\'echet differentiable (Serfling, 1980) with respect to the measure $M$ and $y.$ Let $\Gamma$ be  the Jacobian operator of $T$ with respect to $y$ and given by (Gervini, 2008) 

\begin{eqnarray}
\Gamma=\sum_{k\in U}\frac{1}{||Y_k-m_N||}\left[\mathbf{I}-\frac{(Y_k-m_N)\otimes(Y_k-m_N)}{||Y_k-m_N||^2}\right],
\end{eqnarray}
\noindent where $\mathbf I$ is the identity operator defined by $\mathbf{I}y=y$  and the tensor product of two elements $a$ and $b$ of $L^2[0,T]$ is the rank one operator such that $a\otimes b(y)=<a,y>b$ for all $y\in L^2[0,T].$ One can easily obtain that $\Gamma$ is a strictly positive operator, namely $<\Gamma y,y> >0$ and supposing that $N^{-1}\sum_{k\in U}||Y_k-m_N||^{-1}<\infty, $ we can get following the same arguments as in Cardot \textit{et al.} (2011), that $\Gamma/N$ is a bounded operator, namely  $||\Gamma/N||_{\infty}<\infty$ with $||\Gamma||_{\infty}=\mbox{sup}_{||y||\leq 1}||\Gamma y||. $ We recall that for the operator $\Gamma: L^2[0,T]\longrightarrow L^2[0,T],$  we have 
$$
\Gamma y=\sum_{k\in U}\frac{1}{||Y_k-m_N||}\left(y-\frac{<Y_k-m_N,y>}{||Y_k-m_N||^2}(Y_k-m_N)\right) \quad \mbox{for all}\quad y\in L^2[0,T]
$$
which gives
\begin{eqnarray}
\Gamma y(t)=\sum_{k\in U}\frac{y(t)}{||Y_k-m_N||}-\int_0^T\gamma(r,t)y(r)dr\label{gammaint}
\end{eqnarray}
where 
$$\gamma(r,t)=\sum_{k\in U}\frac{(Y_k(r)-m_N(r))(Y_k(t)-m_N(t))}{||Y_k-m_N||^3}.$$
 The median is defined by the implicit equation (\ref{fonctionalmedian}) and  using then  the implicit function theorem, we obtain that  it exists a unique functional $\widetilde T$ such that  
$$
\widetilde T(M)=m_N
$$ 
and 
$$
\widetilde T(\widehat M)=\widehat m_n.
$$
Moreover, the functional $\widetilde T$ is also Fr\'echet differentiable with respect to $M$ and the derivative of $\widetilde T$ with respect to $M$ is called the influence function and defined, when it exists, as follows
\begin{eqnarray*}
I\widetilde T(M,\xi)=\mbox{lim}_{\lambda\rightarrow 0}\frac{\widetilde T(M+\lambda \delta_{\xi})-\widetilde T(M)}{\lambda}
\end{eqnarray*}
\noindent where $\delta_{\xi}$ is the Dirac function at $\xi \in L^2[0,T]. $ Note that this definition suggested by Deville (1999) and extended to the functional case by Cardot \textit{et al.} (2010) is slightly different from the usual definition of the influence function used in robust statistics (see e.g. Hampel, 1974 or Serfling, 1980), which is based on a probability distribution instead of a finite measure $M. $ A nonstandardized measure $M$ is used in survey sampling because the total mass $N$ may be unknown. \\
Under the asymptotic framework from Deville (1999), we may give a first-order von-Mises (1947) expansion  of $\tilde T$ in $\widehat M/N$ around $M/N,$ 
\begin{eqnarray}
\widetilde T\left(\frac{\widehat M}{N}\right)=\widetilde T\left(\frac{M}{N}\right)+\int I\widetilde T\left(\frac{M}{N},\xi\right)d\left(\frac{\widehat M}{N}-\frac{M}{N}\right)(\xi)+o_p(n^{-1/2})\label{vonmises1}
\end{eqnarray}
which may be written in the equivalent form,
\begin{eqnarray}
\widetilde T(\widehat M)=\widetilde T(M)+\int I\widetilde T\left(M,\xi\right)d(\widehat M-M)(\xi)+o_p(n^{-1/2})\label{vonmises2}
\end{eqnarray}
 since $\widetilde T$  is a functional of degree zero, namely $\widetilde T(M/N)=\widetilde T(M)$ and in this case, $I\widetilde T\left(\frac{M}{N},\xi\right)=N\cdot I\widetilde T\left(M,\xi\right)$ (Deville, 1999).\\ 
\noindent Let $u_k,$ for all $k\in U,$ be the linearized variables    of $\widetilde T(M)=m_N$ and defined as the value of the influence function  $I\widetilde T$ at $\xi=Y_k,$ namely
\begin{eqnarray}
u_k & = & I\widetilde T(M,Y_k)=   \Gamma^{-1} \left( \frac{Y_k-m_N}{||Y_k-m_N||}\right).\label{varlin}
\end{eqnarray}
We have used here the fact that for fixed $y,$ the functional $T(M;y)=-\displaystyle\sum_U\frac{Y_k- y}{||Y_k- y||}$ is a finite population total with influence function at $Y_k$ given by $IT(M,Y_k;y)=-(Y_k-y)/||Y_k-y||$ (Deville, 1999). From the Riesz's theorem, we have that for all bounded $h\in L^2[0,T] $ there is a unique  $f\in L^2[0,T] $ such that $\Gamma f=h$ and $\Gamma f(g)=<h,g>$ for all $g\in L^2[0,T]. $ This unique $f$ will denote $\Gamma^{-1}h$ for a given $h\in L^2[0,T]. $

Hence, the expansion (\ref{vonmises2}) becomes
\begin{eqnarray}
\widehat m_n & = &  m_N +\sum_{k\in s} \frac{u_k}{\pi_k}- \sum_{k\in U} u_k+ o_p(n^{-1/2}).\label{medlin}
\end{eqnarray} 
\noindent The above formula  shows that the nonlinear estimator $\widehat m_n$ may be approximated by the Horvitz-Thompson estimator for the total of the linearized variables $u_k. $ In this way, $u_k$ is  an artificial variable used to compute the approximative variance of $\widehat m_n. $ Now, the linearized variable $u_k$ is also a functional defined on $L^2[0,T]$ and it is unknown since $m_N$ and $\Gamma$ are unknown. We suggest estimating $u_k$ by 
\begin{eqnarray}
\hat u_k=\widehat {\Gamma}^{-1}\left(\frac{Y_k-\widehat{m}_n}{||Y_k-\widehat{m}_n||}\right),\label{estimvarlin}
\end{eqnarray}  
where $\widehat {\Gamma}$ is given by
\begin{eqnarray}
\widehat {\Gamma}=\sum_{k\in s}\frac{1}{\pi_k||Y_k-\widehat{m}_n||}\left[\mathbf{I}-\frac{(Y_k-\widehat{m}_n)\otimes(Y_k-\widehat{m}_n)}{||Y_k-\widehat{m}_n||^2}\right].\label{estimgamma}
\end{eqnarray}

\noindent Using relation (\ref{medlin}), one can obtain the asymptotic variance function of $\widehat{m}_n$ calculated under the sampling design, 
\begin{eqnarray}
var(t)=\sum_{k\in U}\sum_{k\in U}(\pi_{kl}-\pi_k\pi_l)\cdot\frac{u_k(t)}{\pi_k}\cdot\frac{u_l(t)}{\pi_l}=\mathbf{u}'(t)\mathbf{\Delta}\mathbf{u}(t)\quad \mbox{for all}\quad t\in [0,T]\label{variancelin}
\end{eqnarray}
where $\mathbf{u}(t)=(u_k(t))_{k\in U}$ with $u_k(t)$ is given by (\ref{varlin}) and  $\displaystyle\mathbf{\Delta}=\left(\frac{\pi_{kl}-\pi_k\pi_l}{\pi_k\pi_l}\right)_{k,l\in U}$. The variance is estimated by   
\begin{eqnarray}
\widehat{var}(t)=\sum_{k\in s}\sum_{k\in s}\frac{\pi_{kl}-\pi_k\pi_l}{\pi_{kl}}\cdot\frac{\hat u_k(t)}{\pi_k}\cdot\frac{\hat u_l(t)}{\pi_l}=\mathbf{\widehat{u}}'_s(t)\widehat{\mathbf{\Delta}}\mathbf{\widehat{u}}_s(t),\label{varestimt}
\end{eqnarray}
where $\mathbf{\widehat{u}}_s(t)=(\hat u_k(t))_{k\in s}$ with $\hat u_k$ given by (\ref{estimvarlin}) and $\displaystyle\widehat{\mathbf{\Delta}}=\left(\frac{\pi_{kl}-\pi_k\pi_l}{\pi_k\pi_l\pi_{kl}}\right)_{k,l\in s}$.\\

\noindent\textbf{Remark 1}: It is worth mentioning that the linearized variable  $u_k$ plays a central role for the estimation of the median. More exactly, the efficiency of any sampling design used for estimating the median curve depends on  how well it estimates the total of the linearized variable $u_k. $ For example, a stratified strategy will be efficient if the strata are homogeneous with respect to $u_k$ as it will be showed below. Nevertheless, to put in practice such a design requires knowing all $u_k$ which may not be readily available. \\

\noindent\textbf{Remark 2}: In practice, we observe the curves $Y_k$ at $D$ discretized points, say $0\leq t_1< t_2< \ldots< t_D\leq T$ that we suppose to be the same for all the curves. When the discretization points vary to one curve to another, methods described in Ramsay and Silverman (2005) may be employed. In order to compute numerical approximations to integrals and inner products, quadrature rules are used. 

\noindent With discretized points, curves may be viewed as multidimensional vectors, in our case, ${\bf Y}'_k = \left(Y_k(t_1), \dots, Y_k(t_D)\right)$ and  $\mathbf{\widehat{u}}'_k=(\hat u_k(t_1), \ldots, \hat u_k(t_D)). $ 
For each $k\in s,$ we need to compute the estimated linearized variable in points $t_1, \ldots, t_D.$  Let  $\mathbf{\widehat{u}}_s=(\mathbf{\widehat{u}}'_k)_{k\in s}$  be the sample vector of estimated linearized variables which can be derived by solving the $D\times n$ dimensional system 
$$
\widehat{\Gamma} \mathbf{\widehat{u}}'_s=\left(\frac{{\bf Y}_1-\widehat{m}_n}{||{\bf Y}_1-\widehat{m}_n||}, \ldots, \frac{{\bf Y}_N-\widehat{m}_n}{||{\bf Y}_N-\widehat{m}_n||}\right),
$$ 
where $\widehat{\Gamma}$  given by  (\ref{estimgamma}) is replaced by a $D\times D$ symmetric matrix. The variance estimator is then derived directly using (\ref{varestimt}).


\section{Application to the EDF load curves}\label{applicationedf}

\subsection{General settings}\label{generalsettings}
The volume of data treated and analyzed by \'Electricit\'e De France is increasing greatly. In fact, in the next few years Electricit\'e De France plans to install millions of smart electricity meters that will be able to send, on request, electricity consumption measurements every second. Obviously, it will be difficult to store and analyse online all these information. The statistician's challenge is to find a strategy, meaning indicators and estimation methods, capable to give a good description of data and  to used it for forecasting. While working with huge data, methods not being time-consuming are highly desirable.

Our proposal consist in considering the median curve as a robust  indicator of the data and estimating it with probability sampling designs. As Lohr stated in "Sampling: Design and Analysis" (1999): \textit{If a probability sampling design is implemented well, an investigator can use a relatively small sample to make inferences about an arbitrarily large population.}  

Let $U$ be a population of $N=18902$ electricity meters installed in small and large companies sending every 30 minutes the electricity consumption during a period of two weeks. We aim at estimating the median curve of the electricity consumption during the second week whereas  the consumption recorded during the first week will be used as auxiliary information. This means that we have $336$ time measures. So, our study population of curves is a set of $N=18902$ vectors ${\bf Y}'_k = \left(Y_k(t_1), \dots, Y_k(t_D)\right)$ with $D=336.$  Let $X_k$ be the consumption curve for the $k$th firm and recorded  during the first week. The consumption curves present low peaks corresponding to night time measurements and high peaks corresponding to middle day measurements. The electricity consumption decreases roughly around the 250th time measurement which corresponds to the beginning of weekend time. The mean and median curves present the same effect as we can see in Figure \ref{meanmed}.

 We consider  several strategies of fixed size $n=2000$ and we compare them through simulations. We distinguish two kinds of sampling designs whether they use or do not use auxiliary information. If auxiliary information is used at the sampling stage, some changing are needed because the variables involved now are curves. On the opposite situation, the selection of the sample is realized from the sampling frame list as for classical multivariate surveys. Finally, the frame list of French firms is well-constructed being very often updated and most of the designs considered below are usually used in practice.     

\begin{enumerate}
\item\noindent\textit{Simple random sampling without replacement (SRSWOR).}\\

\noindent The SRSWOR sampling is a very simple design easy to put into practice. Every possible subset of  $n$ units in the population has the same chance to be the sample. In a functional framework, the selection of a sample of $n=2000$ curves is performed as for the multivariate surveys, namely $n$ labels are drawn from the list of $N$ companies. The estimation of the median curve with  SRSWOR is obtained from equation (\ref{medestim}) for $\pi_k=n/N,$ namely $\widehat m_n$ is the unique solution of the following equation
 \begin{eqnarray}
 \sum_{k\in s}\frac{Y_k-\widehat m_n}{||Y_k-\widehat m_n||}=0.\label{medianSRS}
 \end{eqnarray} 

The asymptotic variance function is equal to 
$$
var_{SRSWOR}(t)=\displaystyle N^2\left(\frac{1}{n}-\frac{1}{N}\right)S^2_{u(t),U}, \quad\mbox{for all t} 
$$ 
where $S^2_{u(t),U}=\sum_{k\in U}(u_k(t)-\overline u(t))^2/(N-1)$ with $\overline u(t)=\sum_{k\in U} u_k(t)/N. $ This variance  is estimated  by 
\begin{eqnarray}
\widehat{var}_{SRSWOR}(t)=\displaystyle N^2\left(\frac{1}{n}-\frac{1}{N}\right)S^2_{\hat {u}(t),s}, \quad\mbox{for all t}\label{estimvarSRS} 
\end{eqnarray}
where $S^2_{\hat u(t),s}=\sum_{k\in s}(\hat u_k(t)-\overline {\hat u}_s(t))^2/(n-1)$ with $\overline{\hat u}_s(t)=\sum_{k\in s} \hat {u}_k(t)/n$ and $\hat {u}_k(t)$  given by (\ref{estimvarlin}). 


\item\textit{Systematic sampling (SYS).}\\

\noindent We consider the systematic design in its basic form (S\"arndal \textit{et al.}, 1992). The inclusion probabilities are $\pi_k=n/N, $ so the median estimator is obtained according to the same equation (\ref{medianSRS}). It is well-known that the systematic sampling may be very inefficient compared to the SRSWOR sampling if the systematic samples are homogeneous. 
One way to improve the efficiency of SYS sampling is to order the sampling frame list according to an auxiliary variable highly correlated with the variable of interest. In this way, adjacent elements tend to be more similar than elements that are farther apart. In our study, we ordered the frame according to the mean electricity consumption during the first week, namely the variable $\tilde X_k=\sum_{d=1}^D X_k(t_d)/D, $ for all $k\in U$  and $D$ is the number of discretization points in $[0,T]. $ Another trade-off for the simplicity of SYS sampling is  that there is no unbiased estimator of the design variance function since $\pi_{kl}=0$ for all $k$ and $l$ not belonging to the same systematic sample. However, using the ordering according to the variable $\tilde X_k$, the SYS is at least as good as the SRSWOR sampling. So,  we might use the variance estimator appropriate for the SRSWOR design given in (\ref{estimvarSRS}).\\
\noindent Systematic sample is really a special case of cluster sampling, so it is often used when it is difficult to construct a sampling frame in advance.

\item\textit{Stratified sampling with simple random sampling without replacement within strata (STRAT).}\\

\noindent In this case, the population is divided  into $H$ nonoverlapping strata denoted $U_h$ and a simple random sample without replacement is selected independently in each stratum.  Let $n_h$  be the sample size within stratum $h$ and $N_h$ be the stratum size. To obtain  the median estimator, we solve the estimation equation
$$
\sum_{h=1}^H\sum_{s_h}\frac{Y_k-\widehat m_n}{\pi_k^h||Y_k-\widehat m_n||}=0,
$$
where $s_h=s\cap U_h$ and $\pi_k^h=n_h/N_h.$ 
It is well-known that stratification may substantially improve the quality of estimates  compared to simple random sampling without replacement and systematic sampling if the strata are well-constructed. More exactly, the more homogeneous the strata are the more efficient the stratification is. It is worth mentioning that improving the estimation of the median curve  means constructing  strata  homogeneous with respect to the linearized variables, $u_k. $ Indeed, relation (\ref{variancelin}) gives us that the asymptotic variance function of $\widehat m_n$ with STRAT is 
 $$
 var_{STRAT}(t)=\sum_{h=1}^HN_h^2\left(\frac{1}{n_h}-\frac{1}{N_h}\right)S^2_{u(t),U_h},
 $$  
where  $S^2_{u(t),U_h}$ is the population variance within stratum $h$ of $u(t)=(u_k(t))_{k\in U_h}. $ That is, the lower the variation  of the linearized variable within stratum,  the lower the asymptotic variance of $\widehat m_n.$  The variance estimator is the sum of variance estimators (\ref{estimvarSRS}) computed within each stratum. 
 
Usually, one builds the stratification using a variable known on the whole population and strongly correlated  with the  variable of interest. In our case, we suggest two stratification variables computed using the first week: the first one is the linearized variable $u_k$  and the second one is the  consumption $Y_k$  (Cardot and Josserand, 2011). The following two sample allocations are used:
\begin{itemize}
 \item [$\bullet$] the proportional allocation (PROP):  $n_h = n N_h/N$ for all $h=1, \ldots H. $
\item[$\bullet$] the $u^{(1)}$-optimal allocation ($u^1$-OPTIM) as suggested by Cardot and Josserand (2011) and computed here with respect to the variance  $S^2_{u^{(1)}(t),U_h}$ of  the linearized variable computed during the first week and denoted by $u_k^{(1)},$ 
$$
n_h = n \frac{N_h \sqrt{\int_0^TS^2_{u^{(1)}(t),U_h} dt}}{\sum_{h=1}^H N_h \sqrt{\int_0^T S^2_{u^{(1)}(t),U_h} dt}}\quad h=1, \dots, H.
$$

The $u^{(1)}$-optimal allocation is similar to the Neyman optimal allocation but computed using $u^{(1)}$ instead of  $u. $  The $x$-optimal allocation is obtained when the consumption during the first week $X_k$ is used.

\end{itemize}

\textit{Stratification based on the linearized variable during the first week}

The proposed strategy can be split into two steps:

{\bf Step 1:} we calculate the linearized variables $u^{(1)}_k$ for all $k\in U$ during the first week. 

{\bf Step 2:} we stratify the population $U$ using the k-means clustering algorithm with the euclidean distance and applied to  the linearized variables $u_k^{(1)} $ for $k\in U.$ According to within cluster variance considerations, we decide to keep $H=4$ different clusters. The strata sizes as well as both the proportional and $u^{(1)}$-optimal allocation  are given in Table\ref{t1}.


\begin{table}[h!]	

\begin{center}
\begin{tabular}{| l | c | c | c | c |}
\hline
Stratum number & 1 & 2 & 3 & 4\\
\hline
Stratum size $N_h$& 6767 &  2420 & 2503 & 7212 \\
\hline
PROP allocation & 716  & 256 & 265 & 763\\
\hline
$u^{(1)}$-OPTIM allocation & 525  & 395 & 428 & 652\\
\hline
\end{tabular}
\end{center}
\caption{Strata sizes, proportional and $u^{(1)}$-optimal allocations when $n=2000.$ }
\label{t1}
\end{table}

We plot in Figure~\ref{mean_U} (a), the mean of $u_k$ computed during the second week and within the $H=4$ strata. Differences among the strata means are noticeable accounting for a significant gain in efficiency if the proportional allocation is used. Now, to better see what kind of consumers the four strata are built of,  we plot in Figure~\ref{mean_U} (b) the mean of the consumption $Y_k. $  We remark that the stratification based on $u_k^{(1)}$ induces a stratification for the consumption curves also. 

\begin{figure}[h!]
\begin{center}

\begin{tabular}{ll}
\includegraphics[height=9cm,width=8cm]{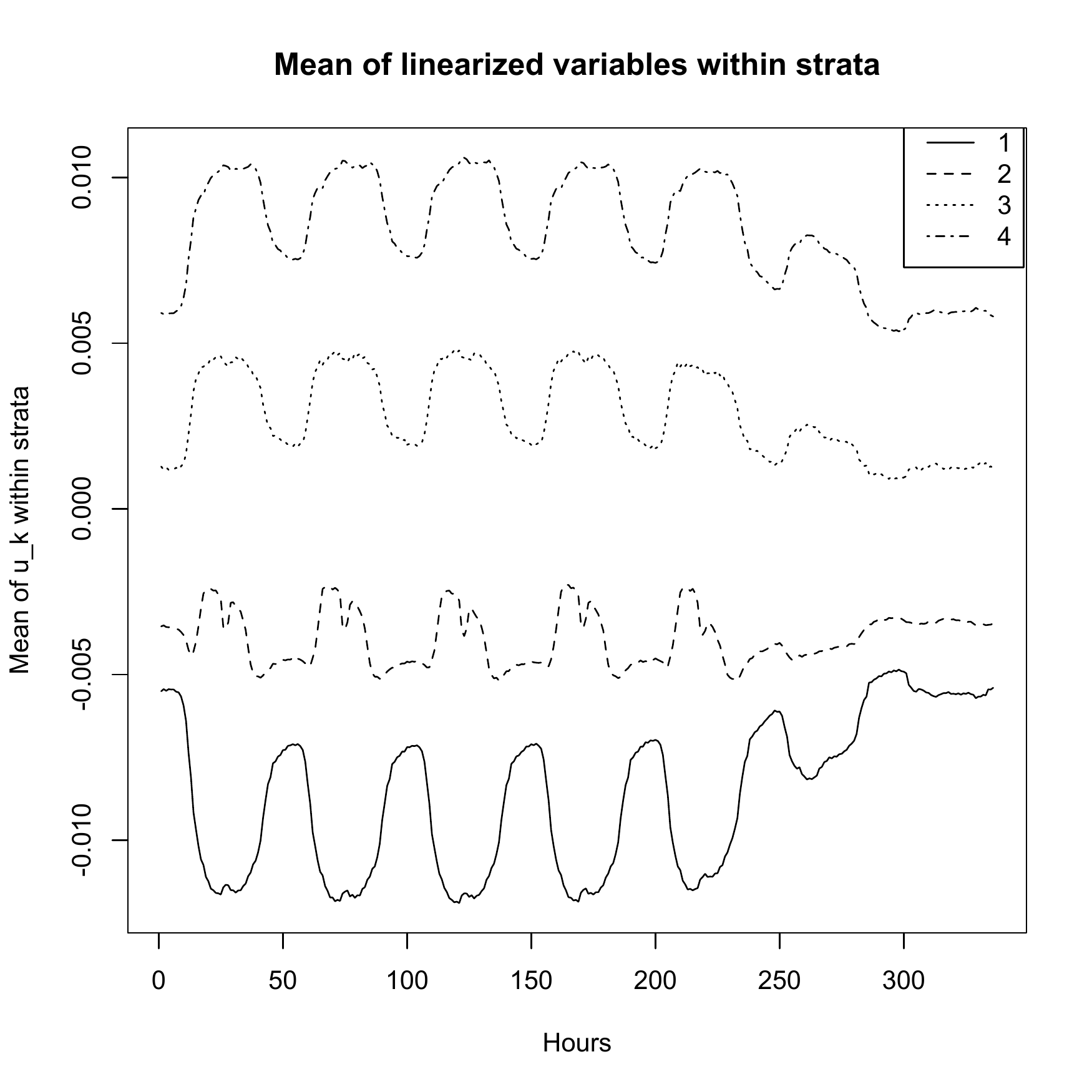} & \includegraphics[height=9cm,width=8cm]{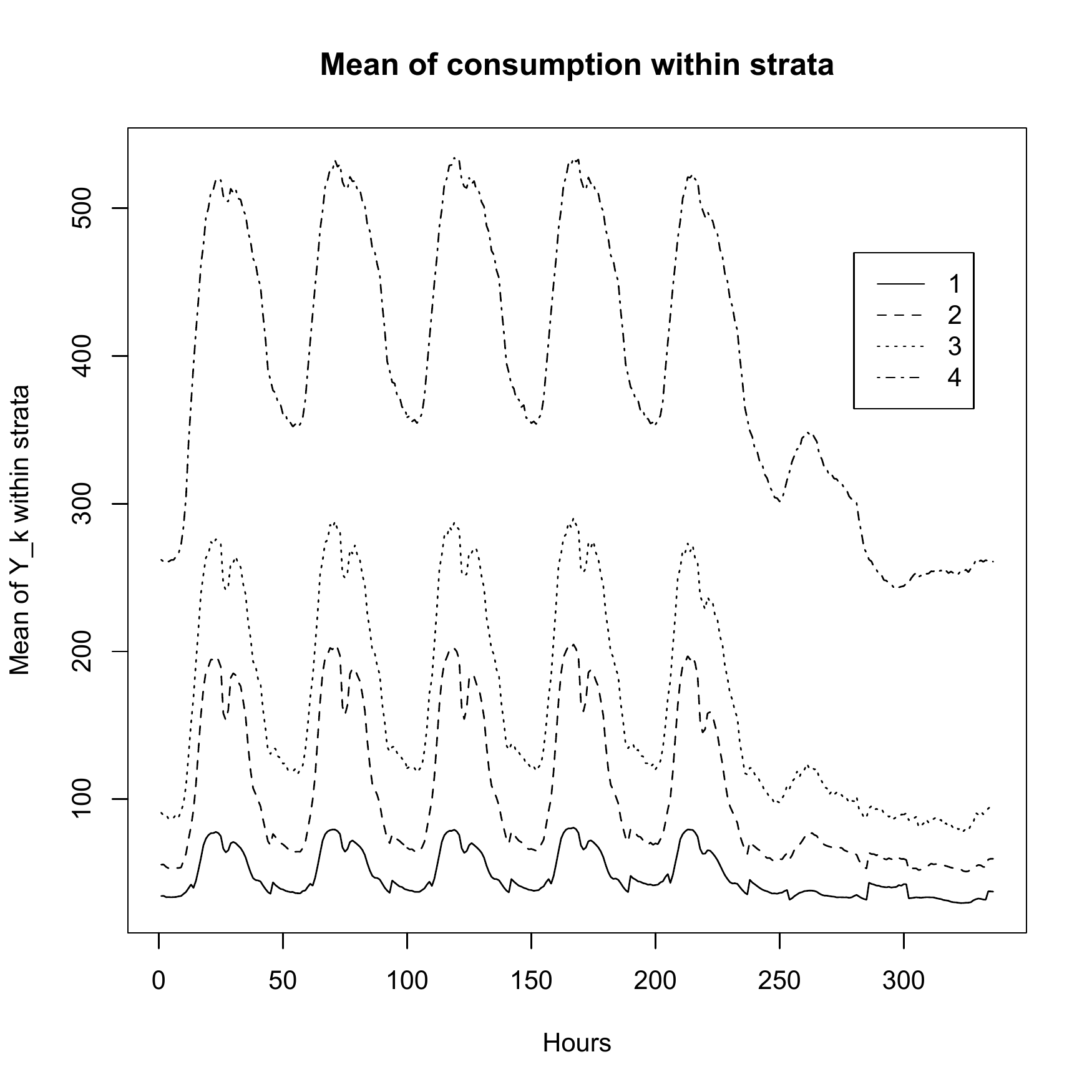}\\ 
\hspace{4cm}{\footnotesize{(a)}} & \hspace{4cm}{\footnotesize{(b)}}\\
\end{tabular}
\end{center}
\caption{Stratification based on the linearized variable: (a) Mean of linearized variables $u_k$ within each stratum. (b) Mean of the consumption curve $Y_k$ within each stratum} 
\label{mean_U}
\end{figure}

\vspace{0.2cm}

\noindent\textit{Stratification based on the consumption curve during the first week}

Cardot and Josserand (2011) suggested taking $H=4$ strata corresponding to the maximum level of consumption during the first week $X_k$ and based on quartiles so that all strata have the same size. We denote the allocation obtained in this way by $x$-OPTIM. The strata sizes as well as both the proportional and the $x$-optimal allocation  are given in Table~\ref{t2}.

\begin{table}[h!]	

\begin{center}
\begin{tabular}{| l | c | c | c | c |}
\hline
Stratum number & 1 & 2 & 3 & 4\\
\hline
Stratum size $N_h$& 4725 &  4726 & 4725 & 4726 \\
\hline
PROP allocation & 500  & 500 & 500 & 500\\
\hline
$x$-OPTIM allocation & 126  & 212 & 333 & 1329\\
\hline
\end{tabular}
\end{center}
\caption{Strata sizes, proportional and $x$-optimal allocations for $n=2000.$ }
\label{t2}
\end{table}

We plot in Figure ~\ref{mean_U2} (b), the consumption mean within strata and during the second week. We notice that the stratum 4 corresponds to consumers with high global levels of consumption, whereas stratum 1, corresponds to consumers with low global of consumption. Figure ~\ref{mean_U2} (a) gives the mean curves of the linearized variable within strata and  computed for the second week. As for the first stratification, the population of the linearized variable curves is also stratified.

\begin{figure}[h!]
\begin{center}

\begin{tabular}{ll}
\includegraphics[height=9cm,width=8cm]{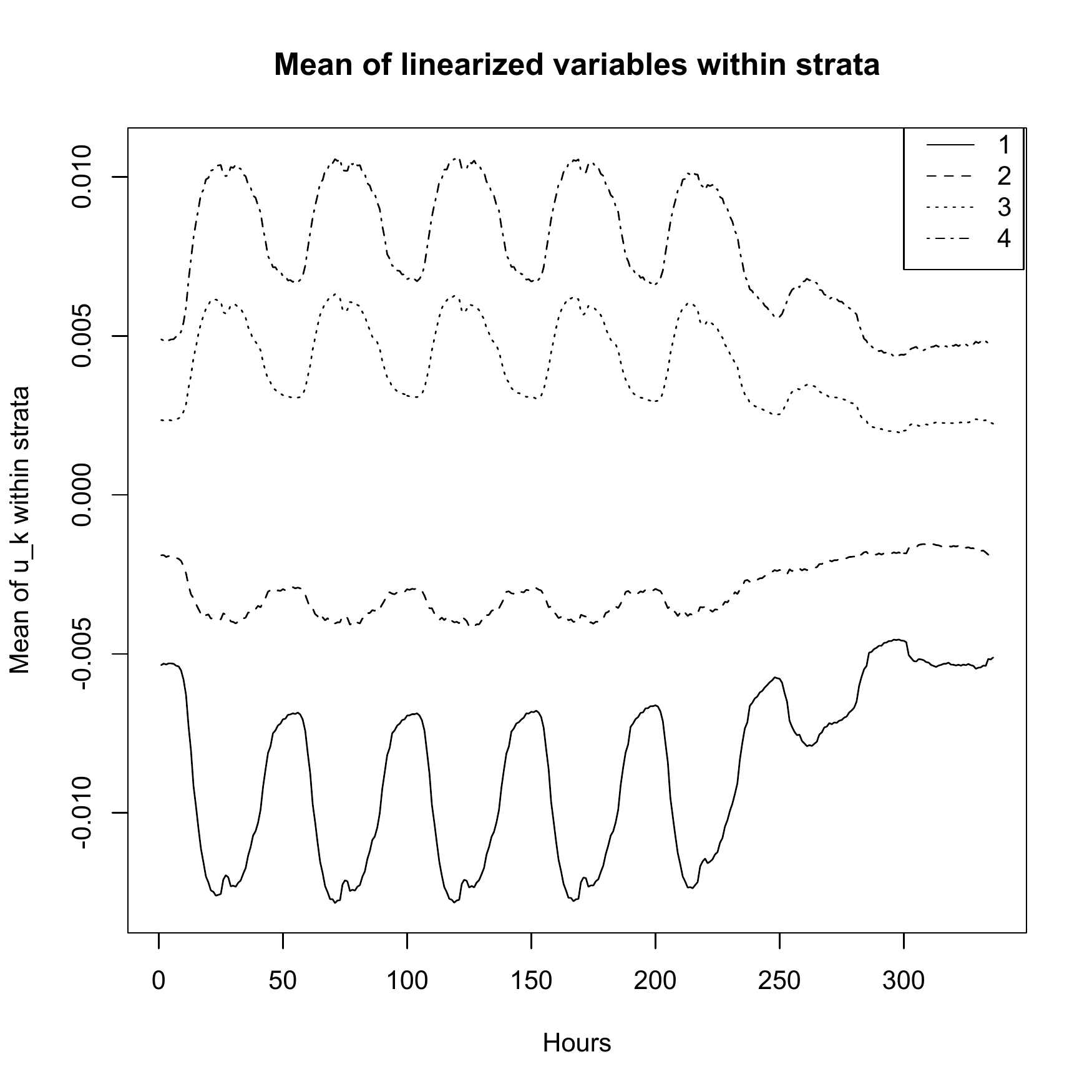} & \includegraphics[height=9cm,width=8cm]{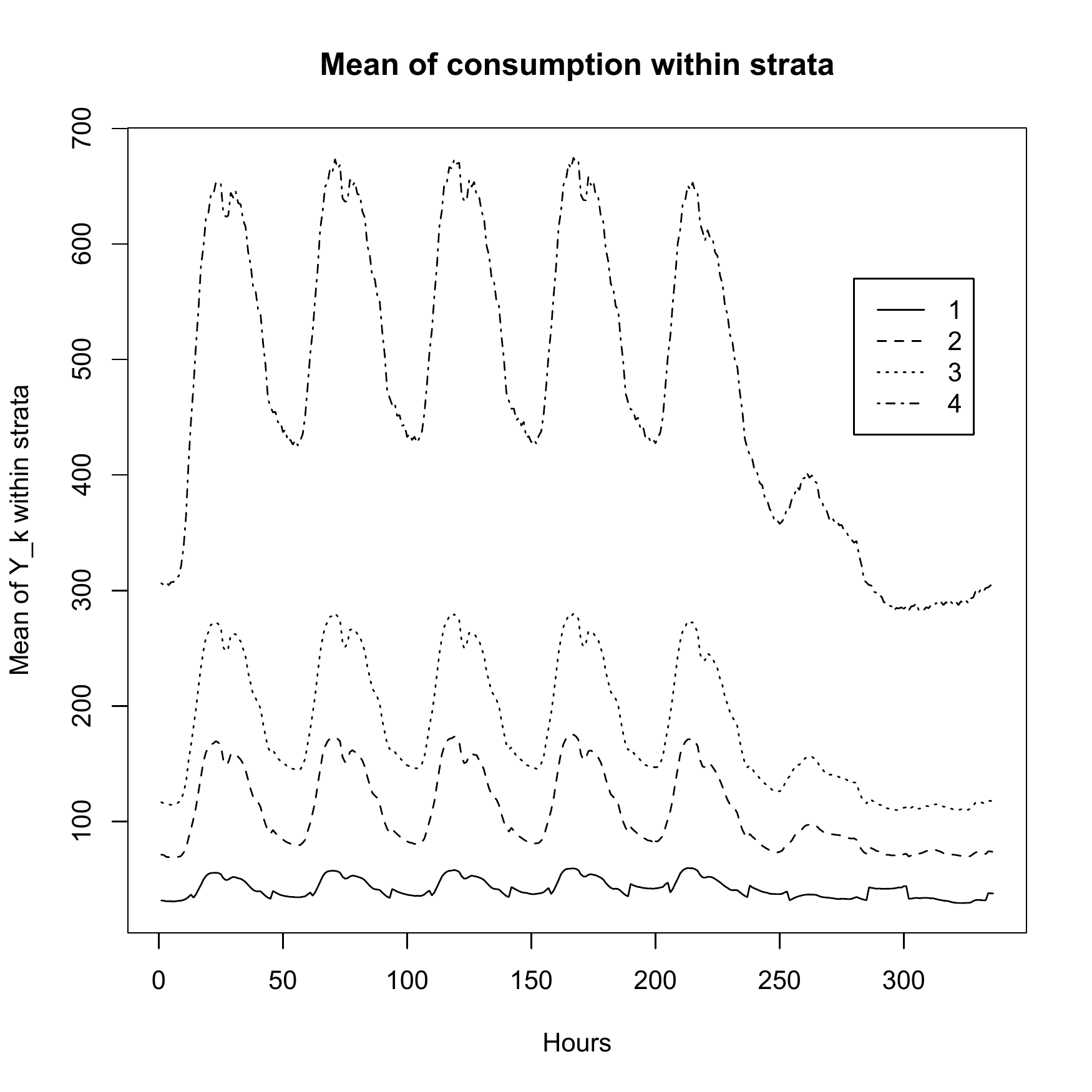}\\ 
\hspace{4cm}{\footnotesize{(a)}} & \hspace{4cm}{\footnotesize{(b)}}\\
\end{tabular}
\caption{Stratification based on the consumption curve: (a) Mean of linearized variables $u_k$ within each stratum. (b) Mean of the consumption curve $Y_k$ within each stratum} 
\end{center}\label{mean_U2}
\end{figure}

\item\textit{Proportional-to-size sampling (PPS)}\\

\noindent Unequal probability designs are widely used in practice because they are usually more efficient than the equal probability designs. In PPS sampling, the sampling is with-replacement and the probability $p_k$ with which the individual $k$ is selected is proportional to a positive measure $X_k,$ where $X_k$ is an auxiliary variable roughly proportional to the study variable $Y_k. $ The probability of selection has the expression
$$
p_k=\frac{X_k}{\sum_{k\in U}X_k}.
$$
In our situation, the study variable is a curve and so  is the auxiliary information. To cope with this problem, we suggest using $p_k$ proportional to the mean of $X_k (t)$ over all $t=1, \ldots, D$ where $D$ is the number of discretization points in the interval $[0,T]. $ This means that 
$$
p_k=\frac{\tilde X_k}{\sum_{k\in U}\tilde X_k},
$$  
where $\tilde X_k=\sum_{t=1}^D X_k(t)/D. $ For our study, we consider again $X_k$ as being  the electricity consumption for the $k$th firm recorded during the first week.  The inclusion probabilities are given by $\pi_k=1-(1-p_k)^n. $ The Horvitz-Thompson  estimator of the median is obtained by solving the equation
\begin{eqnarray}
\sum_{k\in \tilde s}\frac{Y_k-\widehat m_n}{\pi_k||Y_k-\widehat m_n||}=0,\label{ppssampling}
\end{eqnarray}
where $\tilde s$ is the set of distinct elements of $s. $ In with-replacement designs, one may use the Hansen and Hurwitz (1943) estimator which presents the advantage that the variance formula is easier (no double sums are needed). 
\item\textit{Poststratification (POST)}\\

\noindent Let consider now the \textit{poststratification} which is one of the simplest way to take into account auxiliary information in order to improve the Horvitz-Thompson estimator of the median. We suppose that the population is partitioned into subpopulations $U_1, \ldots, U_G$  according to a given classification principle. These subpopulations are called \textit{poststrata} since they do not serve for performing stratified sampling  as described before. Practical considerations may favor some other (perhaps simpler or less costly) designs, such as SRSWOR from the whole population $U.$ After the sample selection, $Y_k$ is observed for the elements $k\in s$ and the sampling frame is used to establish the group each individual belongs to. Remark that group memberships may be unknown before the sample selection which makes impossible to perform the stratified sampling. Nevertheless, the group membership totals $N_g$ are known for all $g=1,\ldots, G$ and this auxiliary information may be used to construct an improved estimator of $m_N. $ The weights used in this case are given by 
$$
w_{ks}=N_g/(\hat N_g\pi_k)\quad \mbox{for all}\quad k\in s_g=s\cap U_g
$$ 
where $\hat N_g=\sum_{k\in s_g}1/\pi_k. $ Hence, the poststratified estimator of $m_N$ is obtained by solving the following equation 
\begin{eqnarray}
\sum_{g=1}^G\sum_{k\in s_g}\frac{N_g}{\hat N_g\pi_k}\frac{Y_k-\widehat m_n}{||Y_k-\widehat m_n||}=0.
\end{eqnarray}

It is important to notice that the weights used here depend on the sample and are more general than the Horvitz-Thompson weights used before in the sense that they include the auxiliary information given by  the group size $N_g.$ In this case, the auxiliary information is used at the estimation stage and not at the design stage. \\
\noindent With SRSWOR sampling from $U,$ the poststratified weights become $w_k=N_g/n_g$ for all $k\in s_g$ and $n_g$ the size of $s_g. $ The median estimator verifies 
 \begin{eqnarray}
\sum_{g=1}^G\sum_{k\in s_g}\frac{N_g}{n_g}\frac{Y_k-\widehat m_n}{||Y_k-\widehat m_n||}=0.\label{poststrat}
\end{eqnarray}

 One should remark that in the case of the poststratification,  the samples sizes $n_g$ are random. The poststratified estimator of the median given by (\ref{poststrat}) cannot be computed if some  sample sizes $n_g$ are equal to zero. However, if the total sample size $n$ is large enough, and if  no group accounts for a very small proportion of the whole population, then the probability of having $n_g=0$ is very small.  S\"arndal \textit{et al.} (1992) suggest aggregating small groups in order to guarantee that $n_g$ are at least 20.  \\
The poststratified estimator is a calibrated estimator (Deville and S\"arndal, 1992)  when the auxiliary information is the group memberships with known totals.  Using the same arguments as in Deville (1999), the expansion given in  (\ref{medlin}) remains valid and as a consequence, the asymptotic variance of $\widehat{m}_n$ is equal to the variance of residuals $E_k=u_k-\overline{u}_g,$ $\overline{u}_g=\sum_{k\in U_g}u_k/N_g$ and with SRSWOR, we obtain
$$
var(t)=N^2\left(\frac{1}{n}-\frac{1}{N}\right)\sum_{g=1}^G\frac{N_g-1}{N-1}S_{u(t),U_g}^2,
$$ 
where $S_{u(t),U_g}^2$ is the group variance. We can remark that the variance agrees very nearly with the asymptotic variance for the stratified sampling with proportional allocation.  
  
We can conclude that the SRSWOR with poststratification is essentially as efficient as STRAT sampling with proportional allocation unless the sample is very small. This result is well-known for multivariate variables $Y_k$ (see e.g., S\"arndal \textit{et al.} 1992).  We have developed above a strategy based on the linearized variables to obtain well-constructed strata. One drawback with that method is that the so-constructed strata reduce the variance for the median estimator but could be inefficient for many other variables. Therefore, using SRSWOR with poststratification will often improve overall efficiency.    


\end{enumerate}

\subsection*{Consistency of $\widehat m_n(t)$ from survey data }

\begin{figure}[h!]
\begin{center}
\includegraphics[height=13cm,width=14cm]{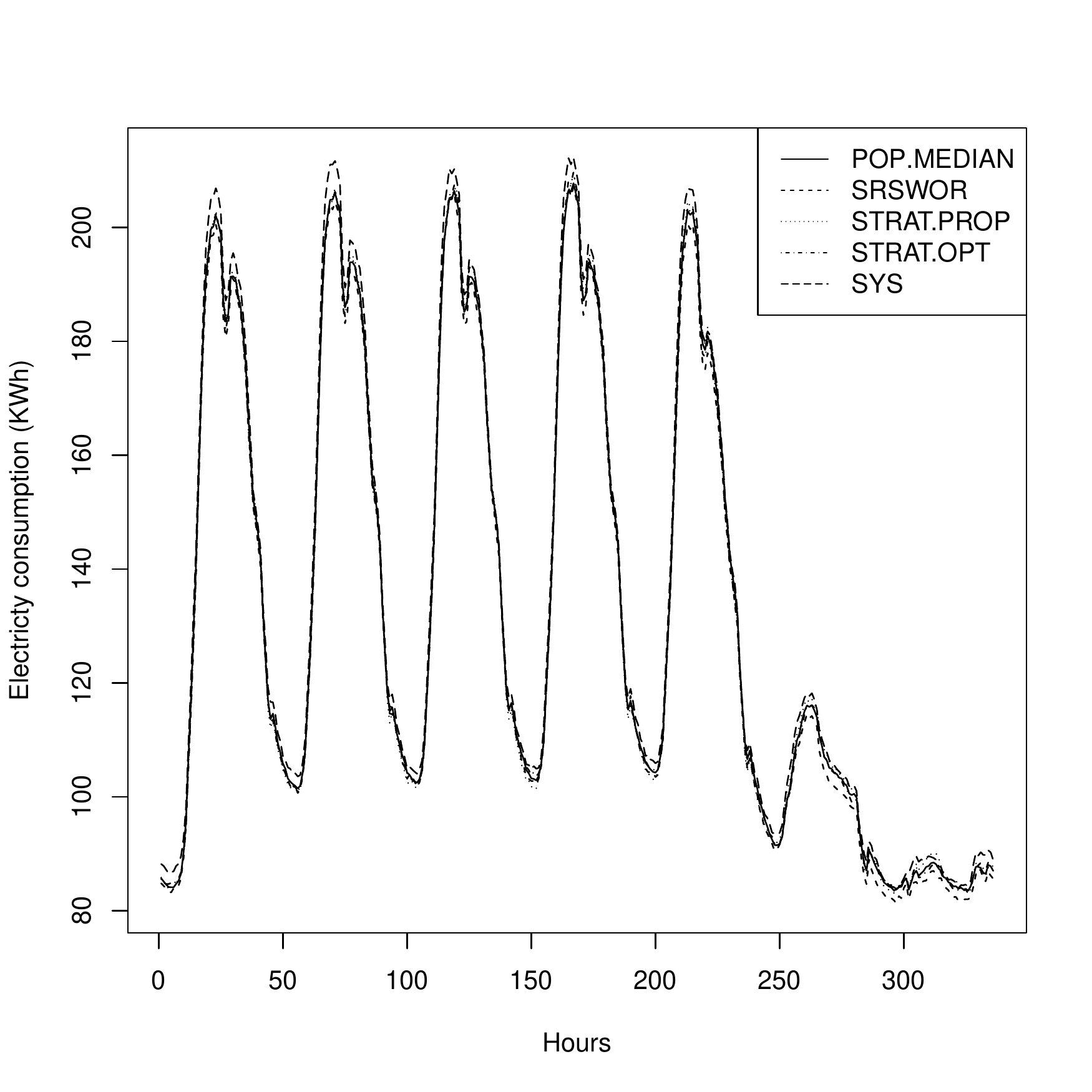} 
\end{center}
\caption{One sample estimation of the median trajectory for $n=2000$ with 4 sampling strategies (SRSWOR, SYS, STRAT.OPT, STRAT.PROP). }
\label{fig.1}
\end{figure}

Focus is now  on the estimation of  the median curve of the consumption recorded during the second week. We consider for that the following sampling designs of size $n=2000$ with the Horvitz-Thompson estimator: SRSWOR, STRAT based on the two stratification variables and with proportional and optimal allocation, SYS, PPS. We consider for our study the POST estimator also.  Figure~(\ref{fig.1}) shows the estimation of the median curve computed from one sample selected according to four sampling designs.

In order to compare these designs, we made 500 replications and considered the following loss criteria,

\begin{eqnarray}
R(\widehat m_n) = \int_0^T |\widehat m_n(t) - m_N(t)|dt.\nonumber
\end{eqnarray}

\begin{figure}
\begin{center}
\includegraphics[height = 15cm,width=15cm]{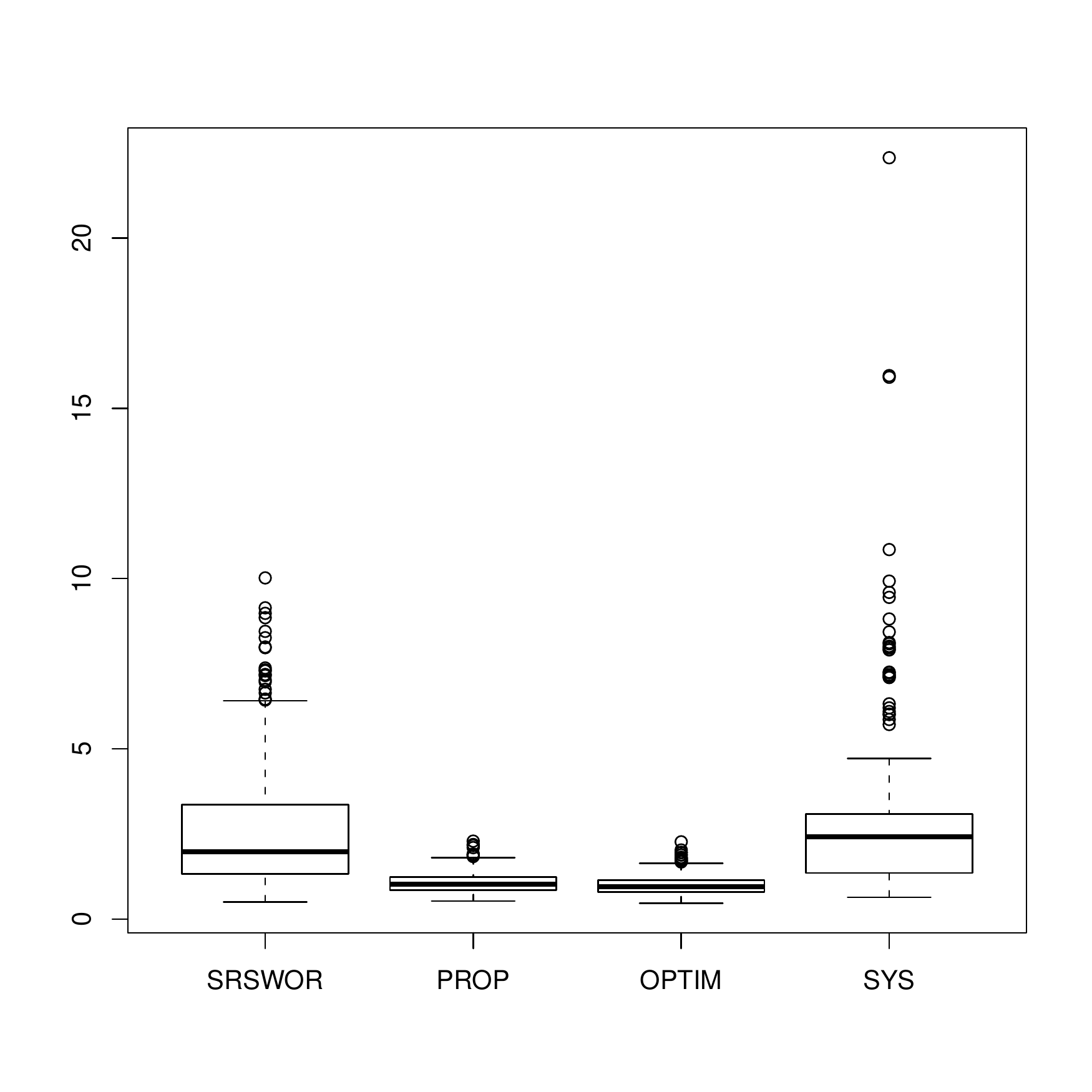}
\end{center}
\caption{Comparison of the distribution of estimation errors of the median curve for SRSWOR, PROP, OPTIM, SYS.}
\label{fig.2}
\end{figure}

\noindent Since in our study we have equally spaced  discretized  time measurements, the above loss criterion is approximated due to quadrature rules by $\frac{1}{D}\sum_{d=1}^D |\widehat m_n(t_d) - m_N(t_d)|.$ Basic statistics, respectively boxplots, for the estimation errors of the median function estimator are given in Table \ref{t3}, respectively Figure \ref{fig.2}. The stratification variable used here is the one based on the linearized variables $u_k^{(1)}. $ First, we can observe that clustering the space of functions by performing stratified sampling leads to an important gain in terms of accuracy of the estimators, dividing by at least  two times the mean error compared to simple random sampling without replacement. We note that the poststratification gives  results similar to those given by stratified sampling with proportional allocation and that the SYS design with ordering on the first week mean consumption is almost as good as the SRSWOR design. A rather surprising result is obtained with PPS sampling. Simulations results not reported here show that this design performs very well for estimating the mean consumption curve, being as good as the stratified sampling but fails for the median curve.  We  believe that this fact is due to numerical  problems  encountered in the resolution of the implicit equation  (\ref{ppssampling}) when a large number of small probabilities of selection $p_k$ are used to estimate $m_N. $ More research is needed to better clarify this issue and to find a way to improve it.

\begin{table}[h!]	

\begin{center}
\begin{tabular}{| l | c | c | c | c |}
\hline
 & Mean & $1^{st}$ quartile & median & $3^{rd}$ quartile\\
\hline
SRSWOR & 2.531 & 1.322 & 1.982 &3.351 \\
\hline
SYS & 2.625 & 1.355 & 2.412 & 3.087 \\
\hline
PROP & 1.060 & 0.8498 & 1.017 & 1.234 \\
\hline
$u^{(1)}$-OPTIM & {\bf 1.000} & {\bf 0.7946} & {\bf 0.9552} & {\bf 1.142} \\
\hline
POST & 1.041 & 0.8275 & 0.9785 & 1.203\\
\hline
PPS & 7.1410 & 2.7880 & 6.1370 & 9.5600\\
\hline
\end{tabular}
\end{center}
\caption{Estimation errors for $m_N$. }
\label{t3}
\end{table}

\newpage
\noindent We also performed simulations for the second stratification based on the first week consumption $X_k$ and the results are given in Table~\ref{tstrat2}.\\
\begin{table}[h!]	
 \begin{center}
\begin{tabular}{| l | c | c | c | c |}
\hline
 & Mean & $1^{st}$ quartile & median & $3^{rd}$ quartile\\
\hline
PROP & 1.7370 & 1.0470  & 1.4860 & 2.2480\\
\hline
$x$-OPTIM & 2.2940 & 1.4660 & 1.9790 &  2.7830\\
\hline
\end{tabular}
\end{center}
\caption{Estimation errors for $m_N$ with stratification based on $Y_k$. }
\label{tstrat2}
\end{table}

\noindent We can remark that STRAT with proportional allocation and stratification based on $u^{(1)}$ gives better results than STRAT with $x$-optimal allocation  stratification based on $X. $ This result is not surprising since in the latter case, the strata have been constructed taken into account the consumption variable and the optimal allocation  has been computed by minimizing the variance for the mean estimator while we are interested here in estimating the median curve. This is why, the proportional allocation is usually advisable  with multipurpose surveys. Moreover, if we compare the two stratifications, we remark that the stratification based on the consumption variable is less  efficient than the stratification based on the linearized variable but it remains still better than the SRSWOR or SYS designs. \\

\noindent Both stratifications used in this paper need the consumption curve $Y_k$ computed during the first week for all the individuals from the population. Sometimes, this can be too costly to obtain or even impossible because of storage or confidentiality constraints. In such situations, some other  stratification variables may be considered such as for example, the electricity power given by the subscribed contract between  one firm and EDF.



\subsection*{Consistency of the variance function estimation from survey data}


\noindent We analyze in the following the estimator for the variance function $var(t)$ when the SRSWOR and STRAT designs are used.  To judge the quality of the estimators,  we use the following criterion

\begin{eqnarray}
R(\widehat{var}) = \int_0^T |\widehat {var}(t) - var(t)| \;dt.\nonumber
\end{eqnarray}

\begin{figure}
\begin{center}
\includegraphics[height=9cm,width=10cm]{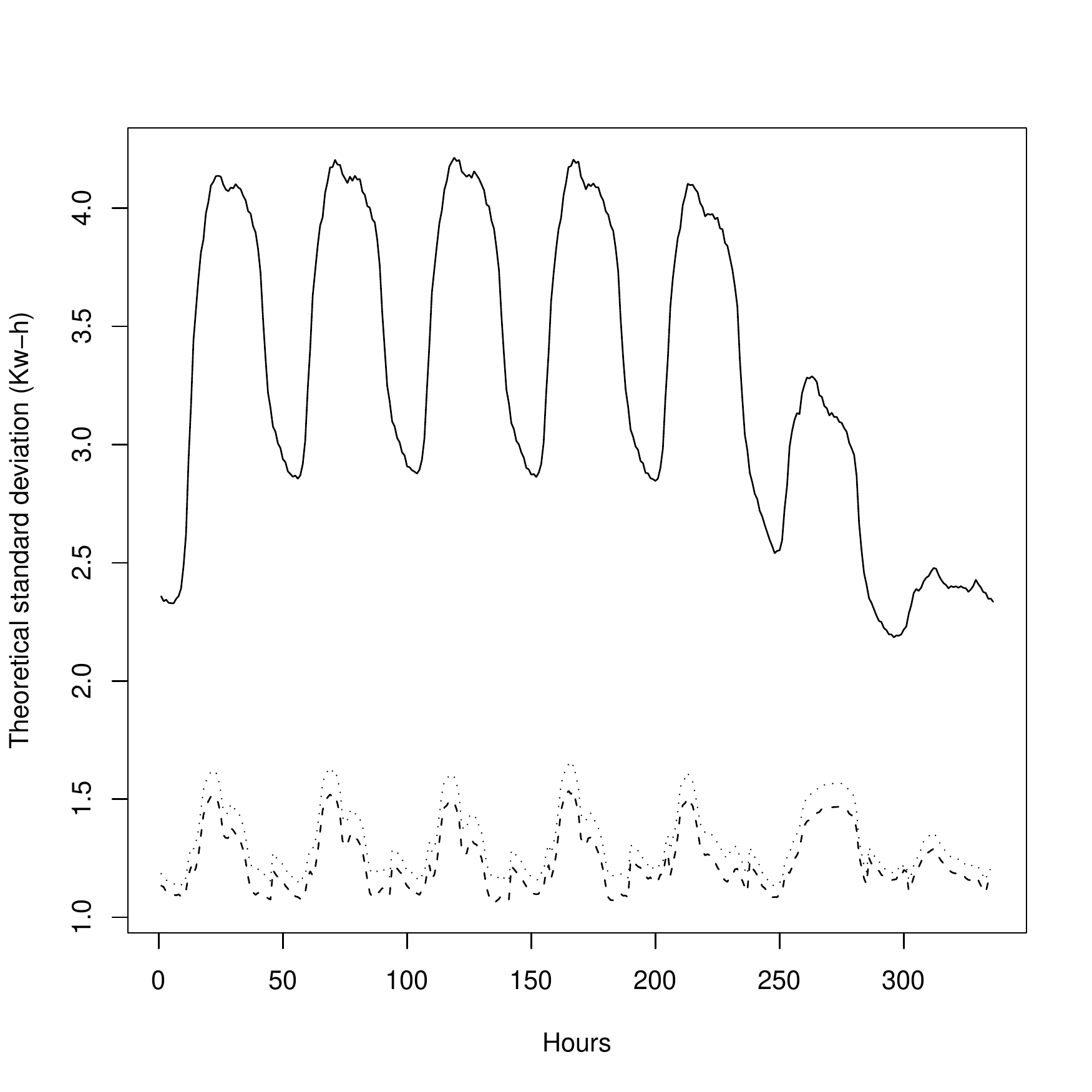}
\end{center}
\caption{Theoretical standard deviation function of $\widehat m_n(t)$ for simple random sampling without replacement (solid line), stratified sampling with proportional allocation (dotted line) and stratified sampling with $u^{(1)}$-optimal allocation (dashed line).}
\label{troisvar}
\end{figure}

\noindent We give in Table \ref{t4} statistics for the estimation errors of the  variance function estimation with SRSWOR and stratified sampling with proportional and $u^{(1)}$-optimal allocations. Figure~\ref{troisvar} gives the theoretical standard deviation function curves of $\widehat m_n$, $\sqrt{var(t)}$, with the considered designs. 
\begin{table}[h!]	

\begin{center}
\begin{tabular}{| l | c | c | c | c |}
\hline
 & Mean & $1^{st}$ quartile & median & $3^{rd}$ quartile\\
\hline
SRSWOR & 0.599 & 0.339 & 0.506 & 0.750 \\
\hline
PROP & 0.068 & 0.055 & 0.064 & 0.076 \\
\hline
$u^{(1)}$-OPTIM & {\bf 0.056} & {\bf 0.047} & {\bf 0.053} & {\bf 0.062} \\
\hline
\end{tabular}
\end{center}
\caption{Statistics about the estimation errors for $var(t)$.}
\label{t4}
\end{table}

One can remark that the theoretical variance is much smaller, at all instants $t$, for the stratified sampling with optimal allocation rule. The stratified sampling with optimal allocation gives more accurate estimation of $var(t)$ than the other strategies. We can observe that clustering the space of functions by performing stratified sampling may leads to a considerable gain in terms of accuracy of the estimators of the variance function, dividing by ten the mean error compared to simple random sampling without replacement. Moreover, there is also a  difference between proportional and optimal allocations rules, for example the third quartile in optimal case is lower than the median loss in the proportional case.

\section{Conclusion and perspectives}
In this paper, we have developed  a survey sampling approach for estimating the median of a functional variable.  From a practical point of view, an appealing consequence of the new methodology is that the proposed estimators are faster to calculate. 
The experimental results on a test population of electricity consumption curves confirm that even with high dimensional data, stratification associated with the optimal allocation rule leads to important reduction  of the variance estimators. Having  appropriate strata is the key for getting more accurate estimators and the k-means algorithm is well adapted in this situation. 
Nevertheless, choosing the stratification variables is a rather complex issue and more work is needed in this direction.



A challenging future research avenue concerns the use of auxiliary information at the estimation stage. While, in this paper, we have concentrated on the estimation of the median using the Horvitz-Thompson estimator or the poststratified estimator,  more complex estimators using functional regression models can be developed. For example, it is possible to set a linear functional model which explains the functional variable $Y_k$ using  a scalar $X_k$ and to develop a regression estimator for the median curve. Developing a general framework for regression estimators for the median curve is left for future studies. 

\vspace{0.3cm}
\noindent \textbf{Acknowledgments}\\
The authors thank the two anonymous referees, and the associate editor for their constructive remarks that helped to improve the manuscript.

\section*{Bibliography}
%
\begin{description}
\item Brown, B.M. (1983)
\newblock Statistical Use of the Spatial Median,
\newblock{\em Journal of
the Royal Statistical Society}, B, \textbf{45}, 25-30.

\item Cadre, B. (2001). 
\newblock Convergent estimators for the $L^1$-median of a Banach valued random variable. \textit{Statistics}, \textbf{35} (4), 509-521. 
\item Cardot, H., C\'enac, P. and Zitt, P.-A. (2011). 
\newblock Efficient and fast estimation of the geometric median in Hilbert spaces with an averaged stochastic gradient algorithm. \textit{Bernoulli, to appear}.

\item Cardot, H., Chaouch, M., Goga, C. and Labru\`ere, C. (2010), 
Properties of design-based functional principal components analysis, \textit{Journal of Statistical Planning and Inference}, \textbf{140}, 75-91.

\item Cardot, H. and Josserand, E. (2011), 
\newblock Horvitz-Thompson estimators for functional data: asymptotic confidence bands and optimal allocation for stratified sampling, 
\textit{Biometrika}, \textbf{98}, 107-118.

\item Chaouch, M. and Goga, C. (2010), Design-based estimation for geometric quantile with application to outlier detection, \textit{Computational Statistics and Data Analysis}, \textbf{54}, 2214-2229.

\item Chaudhuri, P. (1996)
\newblock On a Geometric Notion of Quantiles for Multivariate Data, 
\newblock{\em Journal of the American Statistical Association}, \textbf{91}, pp. 862-872.


\item Chiky, R. and H\'ebrail, G. (2008). Summarizing distributed data streams for storage in data warehouses. in DaWaK 2008, I-Y. Song, J. Eder and T. M. Nguyen, eds. \emph{Lecture Notes in Computer Science}, Springer, 65-74.

 \item Deville, J.C. (1999).
\newblock Variance estimation for complex statistics and estimators: linearization and residual techniques.
\newblock {\em Survey Methodology},  \textbf{25}, 193-203.

\item Deville, J. C. and S\"arndal, C. E. (1992).
\newblock Calibration estimators in survey sampling,
\newblock{\em Journal of the American Statistical Association,} \textbf{87}, 376-382.

\item Fuller, W.A. (2009). 
\emph{Sampling Statistics}. John Wiley and Sons.

\item Gervini, D. (2008).
\newblock Robust functional estimation using the spatial median and
spherical principal components.
\newblock {\em Biometrika}, {\bf 95}, 587-600.

\item Gini, K. and Galvani, L. (1929).
\newblock Di talune estensioni dei concetti di media ai caratteri qualitativi.
\newblock {\em Metron}, {\bf 8}, 3-209.

\item Goga, C. and Ruiz-Gazen, A. (2011)
\newblock{Efficient  Estimation of Nonlinear Finite Population Parameters Using Nonparametrics},
\newblock{submitted}.

\item Gower, J.C. (1974). Algorithm as 78: The mediancentre. \textit{Journal of the Royal Statistical Society, Series C, Applied Statistics}, 23, 466-470.
\item Haldane, J.B.S. (1948). 
\newblock Note on the median of a multivariate distribution. \textit{Biometrika}, \textbf{35}, 414-417.
 
\item Hayford, J. F. (1902).
\newblock What is the center of an area, or the center of a population ? 
\newblock {\em Journal of the American Statistical Association}, {\bf 8}, 47-58.

\item Hampel, F.R. (1974).
\newblock The influence curve and its role in robust statistics.
\newblock{\em Journal of the American Statistical Association}, \textbf{69}, 383-393.

\item Hansen, M. H. and Hurwitz, W.N. (1943).
\newblock On the theory of sampling from finite population
\newblock {\em Annals of Mathematical Statistics}, \textbf{14}, 333-362.

\item Horvitz, D.G. and Thompson, D.J. (1952), 
\newblock A generalization of sampling without replacement from a finite universe, 
\newblock{\em Journal of the American Statistical Association}, \textbf{47}, 663-685.

\item Kemperman, J.H.B. (1987), 
\newblock The median of a finite measure on a Banach space, 
\newblock{\em In: Dodge, Y. (Ed.), Statistical Data Analysis Based on the $L_1$ Norm and Related Methods, North-Holland, Amesterdam},  217-230.

\item Korn, E.L.  and Graubard, B.I. (1999). \textit{Analysis of Health Surveys}, Wiley, New York.

\item Koenker, R., and Bassett, G. (1978)
\newblock Regression Quantiles,
\newblock{\em Econometrica}, \textbf{46}, 33-50.

\item Lehtonen, R. and Pahkinen, E. (2004). \textit{Practical Methods for Design and Analysis of Complex Surveys}, Wiley, New York.

\item Lohr, S. L. (1999).  \textit{Sampling: Design and Analysis}, Duxbury Press.

\item Ramsay, J.O. and Silverman, B.W. (2005),
\newblock{\em Functional Data Analysis}, 2nd edition, Springer, Berlin.

\item S\"arndal, C.E., Swensson, B. and Wretman, J. (1992). \textit{Model Assisted Survey Sampling}. Springer-Verlag. 

\item Serfling, R. (1980).
\textit{Approximation Theorems of Mathematical Statistics}, Wiley, New York.

\item Serfling, R. (2002)
\newblock{ Quantile functions for multivariate analysis: approaches and applications}. 
\newblock{\em Statistica Neerlandica}, 56, 214-232.

\item Small, C.G. (1990). \textit{A survey of multidimensional medians, International Statistical Review}, \textbf{58}, 263-277.

\item Vardi, Y and Zhang, C.H. (2000). \newblock The multivariate $L_1$-median and associated data depth. 
\newblock {\em Proc. Natl. Acad. Sci. USA}, {\bf 97}, 1423-1426.

\item von-Mises, R. (1947). \newblock On the asymptotic distribution of differentiable statistical functions. 
\newblock{\em Annals of Mathematical Statistics}, \textbf{18}, 309-348.

\item Weber, A. (1909), 
\newblock Uber Den Standard Der Industrien, Tubingen. English translation by C. J. Freidrich (1929), 
\newblock{em Alfred Weber's Theory of Location of Industries}, Chicago: Chicago University Press.

\end{description}
\end{document}